# A field guide to cultivating computational biology

Anne E. Carpenter, Imaging Platform, Broad Institute of MIT and Harvard, Cambridge MA USA
https://orcid.org/0000-0003-1555-8261

Casey S. Greene, Center for Health AI, University of Colorado School of Medicine, Aurora, CO, USA
https://orcid.org/0000-0001-8713-9213

Piero Carninci, RIKEN Center for Integrative Medical Sciences Yokohama, Kanagawa, Japan and Human Technopole, Via Belgioioso 171, 20157 Milano MI, Italy. https://orcid.org/0000-0001-7202-7243

Benilton S Carvalho, Department of Statistics, Institute of Mathematics, Statistics and Scientific Computing, University of Campinas, Campinas, Brazil https://orcid.org/0000-0001-5122-5646

Michiel de Hoon, RIKEN Center for Integrative Medical Sciences, Yokohama, Kanagawa, Japan https://orcid.org/0000-0003/0489-2352

Stacey Finley, Department of Biomedical Engineering, Quantitative and Computational Biology, and Chemical Engineering & Materials Science, University of Southern California, USA
https://orcid.org/0000-0001-6901-3692

Kim-Anh Lê Cao, Melbourne Integrative Genomics, School of Mathematics and Statistics, The University of Melbourne, Australia https://orcid.org/0000-0003-3923-1116

Jerry S.H. Lee, Departments of Medicine/Oncology, Chemical Engineering, and Material Sciences, University of Southern California, CA, USA https://orcid.org/0000-0003-1515-0952

Luigi Marchionni, Department of Pathology and Laboratory Medicine, Weill-Cornell Medicine, New York, NY, USA https://orcid.org/0000-0002-7336-8071

Suzanne S. Sindi, Department of Applied Mathematics, University of California, Merced, Merced, CA, USA https://orcid.org/0000-0003-2742-4332

Fabian J. Theis, Institute of Computational Biology, Helmholtz Center Munich and Department of Mathematics, Technical University of Munich, Germany https://orcid.org/0000-0002-2419-1943

Gregory P. Way, Imaging Platform, Broad Institute of MIT and Harvard, Cambridge MA USA
https://orcid.org/0000-0002-0503-9348

Jean Y.H. Yang Charles Perkins Centre and School of Mathematics and Statistics, The University of Sydney, Australia https://orcid.org/0000-0002-5271-2603

Elana J. Fertig*, Convergence Institute, Departments of Oncology, Biomedical Engineering, and Applied Mathematics and Statistics, Johns Hopkins University, Baltimore, MD, USA
https://orcid.org/0000-0003-3204-342X

* corresponding author
ejfertig@jhmi.edu

**Abstract**

Biomedical research centers can empower basic discovery and novel therapeutic strategies by leveraging their large-scale datasets from experiments and patients. This data, together with new technologies to create and analyze it, has ushered in an era of data-driven discovery which requires moving beyond the traditional individual, single-discipline investigator research model. This interdisciplinary niche is where computational biology thrives. It has matured over the past three decades and made major contributions to scientific knowledge and human health, yet researchers in the field often languish in career advancement, publication, and grant review. We propose solutions for individual scientists, institutions, journal publishers, funding agencies, and educators.

A modern research project may include multiple model systems and assay/data types and require complex computational strategies, making it difficult or impossible for one scientist to effectively design or execute all aspects of the project. We see the solution to unleashing the power of biological data as deep integration between biology and computational science (defined extremely broadly in this article as all quantitative approaches including not just computer science but also statistics, mathematics, data science, and so on). Deep integration can come in many flavors (1) a single researcher with expertise in biology and computation, including (2) a computational biology lab with both wet-lab and computational researchers working closely together with scientific leaders from both sides, (3) paired laboratories from each side that commit to working closely together for a project or across the laboratories' lifetimes, (4) computational labs embedded in biological departments, or (5) computational departments that value biomedical applications.

None of these models work well without mutual respect: individual scientists cannot stick strictly to their "home" discipline or treat one as working in service of another. Interdisciplinary collaborations, whether at the individual scientist or lead/chief/principal investigator level, enable the deepest understanding of data, with biologists able to delve into the underlying biological mechanisms of their system and computational scientists being able to design methods to uncover the underlying biology. A new generation of scientists who are truly a hybrid of different fields is arriving on the scene to advance biology.

Visionaries a decade ago aspired to bridge the domains of computational sciences and biology. Since then, computational biology has emerged as a mature scientific discipline, with investigators following new research models (Loman and Watson 2013; Markowetz 2017) and educational training (Form and Lewitter 2011; Ayoob and Kangas 2020; Demharter et al. 2017; Mulder et al. 2018; Carey and Papin 2018). Despite the fact that dramatic advancements have been driven by computational biology, too often researchers choosing this path languish in career advancement, publication, and grant review. Many traditional schemes for publication, evaluation, advancement, and funding were built for the era of single-discipline science and fail to support the interdisciplinary team science necessary for advancing computational biology.

Solving the bottlenecks in data-driven discovery requires empowering computational biologists and rewarding interdisciplinary and team science. We suggest concrete actions for individual scientists, administrators, and funding agencies to cultivate computational biology to advance biomedical science, roughly in the order of a project's life cycle, from design to execution to sharing data and results to assessment, education, and, finally, infrastructure.

**Respect collaborators' specific research interests and motivations**
Computational biology hinges on mutual respect between researchers from different disciplines, and a key element of respect is understanding a colleague's particular expertise. Computationalists do not like to be seen as "just" running the numbers any more than biologists appreciate the perception that they are "just" a pair of hands that produced the data. Statistics, database structures, clinical informatics, genetics, epigenetics, genomics, imaging, single-cell technologies, structure prediction, algorithm development, machine learning, and mechanistic modeling are all distinct fields. Biologists should not be offended if a particular collaboration idea does not fit a computational biologist's research agenda or expertise. Some institutions subsidize a core bioinformatics facility offering services across a spectrum of data types, but their mission, expertise, and/or bandwidth often limits them to standardized analysis pipelines that may miss insights in the data. In contrast, a computational biologist's research laboratory often must focus on novel methods development in a particular area for career advancement.

The current system incentivizes mechanism and translational discovery for biology but methodological advances for computational sciences. This explains a common disconnect when collaborating: projects that require routine use of existing methodology typically provide little benefit to the computational person's academic record no matter how innovative a particular dataset. Transparent and open discussion of each investigator's expertise, limitations, goals, expectations, deliverables, and publication strategies upfront can facilitate productive interdisciplinary collaboration. Matching research interests can facilitate dual submission of methodological and biological manuscripts, which provides leading roles for all investigators in the research team.

**Seek necessary input during project design and throughout the life cycle of the project**
Scientists lacking particular expertise for a project should engage collaborators with such expertise from the beginning of the research project lifecycle ("Integrating Quantitative Approaches in Cancer Research and Oncology" 2021). Computational scientists may have critical insights that impact the scope of the biological questions, study design, analysis, and interpretation. Similarly, biologists' early involvement may influence the algorithmic approach, data visualization, and refinement of analysis. The onset of a project is the ideal time to plot out feasibility, brainstorm solutions, and avoid costly missteps. It is also the time to establish clear communication and define expectations and responsibilities, in particular in the gray area between (experimental) data generation and (computational) data analysis. Computational scientists should learn about data acquisition techniques and the factors that influence data quality, as well as the cost to collect new datasets. As the project progresses, collaborators must understand that data analysis is rarely turnkey: optimal analysis requires iteration and engagement and can yield fruitful discovery and new questions to ask.

**Provide and preserve budgets for computational biologists' work**
There is a common misconception that the lack of physical experimentation and laboratory supplies makes analysis work automated, quick, and inexpensive. This stems from a sense that labor/time is a "free" resource in academia, and perhaps that each analysis should only need to be run a single time. In reality, even for well-established data types, analysis can often take as much or more time and effort to perform as generating the data at the bench. Moreover, it typically also requires pipeline optimization, software development and maintenance, and user interfaces so that methods remain usable beyond the scope of a single publication or project. Given that computational biologists often command higher salaries due to competition with industry, researchers should therefore ensure space in the budget to support this effort as well as computing costs on projects.

Scientists, institutions, and funders should also *preserve* budgets for collaborative researchers. When funding agencies impose budget cuts, collaborators' budgets are often the first to go. This can substantially impact computational laboratories' ability to provide projects and salaries to their members, especially considering that time spent providing preliminary data or ideas for the proposal cannot be recouped. In one computational biology laboratory, one-third of its collaborators' funded grants cut their budget entirely and another third cut their budget partially - by an average of 90% (Carpenter 2018). Although principal investigators are authorized to make budget changes at will, they should consider the impact of doing so on the long-term health of their relationship with their collaborators and the scientific community more generally. Institutions and agencies can help promote good behavior by a simple policy change: by default, budget cuts are distributed evenly; the lead investigator can then propose changes if needed. Societies could advocate for this change at the funding agencies.

**Change publication conventions and perceptions**
Many high-impact papers have computational biologists in key authorship positions (first and last, customarily the most highly-valued spots in biology journals). Nevertheless, middle-author placements are very common for computational biologists, reflecting a methodological contribution on a paper to address a particular biological question. Co-first or co-senior authorships provide a means to provide credit when computational contributions are significant to a paper, but such designations are often dramatically discounted in grant, hiring, promotion, and tenure assessments. We have seen recent comments such as: "No recent first or senior author papers were listed (though one acknowledges several were 'starred' co-first authored)". Although the idea of breaking away from a linear, rank-ordered list of authors may seem unimaginable, we note that the standards of authorship order in biology journals are not universal. In fields such as mathematics and physics, authors are listed alphabetically. This further complicates the evaluation of computational publication records by biologists, who are typically not aware of the relative reputation of computational publishing venues. Other alternatives are possible, and indeed scientific publishing has experienced dramatic change in recent years (Callaway 2020). One option is for journals to formally encourage swapping the order of authors that have been designated as "equally contributing" via their display interfaces. Another might be to allow designations such as the "corresponding author on experimental aspects" and the "corresponding author on bioinformatics".

Fundamentally, these issues will only be solved through educating paper and grant reviewers on how to best understand and appreciate different kinds of scientists' contributions. From calling computational biologists "research parasites" (Longo and Drazen 2016) to "mathematical service units" (Markowetz 2017), it is clear we have a long way to go. One innovative solution was undertaken in 2021 by the Australian Research Council (ARC): a 500-word section was added to proposals for applicants to describe "research context" and "explain the relative importance of different research outputs and the expectations" in the specific discipline/s of the applicants. The São Paulo Research Foundation (FAPESP) updated its biosketch format to include up to ten most relevant scientific results (articles, registered software, patents, and other items that can be chosen at the author's discretion) and verifiable description of the impact of these items (citations, prizes, impact in public policies, and so on). This provides an opportunity to explain author order and other information helpful to assess merit within a discipline.

Another complication to evaluating computational biologists is that their primary research output may not be papers but instead valuable software or data (Perkel 2021). The U.S. National Institutes of Health changed its biosketch format to explicitly encourage software products to be mentioned in its "Contributions to Science" section, where usage and impact can be described. It is becoming more common to publish software and data papers, aided in part by some journals creating article types specifically for software and data. Allowing updated versions of software to be published as well provides academic credit for the largely thankless task of software maintenance over time.

Schemes for tracking the usage of software and data are a work in progress (Wade and Williams 2021). Unfortunately, for software and data, citations of their related papers (a key metric used in major career decisions) are inconsistent. Every scientist can help ensure that software and data creators receive credit for their work by citing the related papers rather than just mentioning the name of the software or database (or omitting mention entirely), which leads to inconsistent tracking when evaluating investigator impact (Singh Chawla 2016). Journals can require that software and data are properly cited.

**Establish academic structures and review panels that reward the team science efforts employed in computational biology**
Wet-lab biologists trained in traditional evaluation schemes can be quick to dismiss a researcher with a lack of a single driving biological question for the laboratory, many middle-authored papers, publication in computational conferences rather than journals, a low citation count or h-index (due to field-specific differences), or funding through grants led by others, leading to comments such as "How do people like you ever get last-author papers?" and "an overly strong reliance on collaborators" (Markowetz 2017). Computational scientists can dismiss a body of work as too applied, with not enough theory or conference papers that are the currency of the field. Evaluation panels should therefore include interdisciplinary researchers and be provided with guidelines about the challenges of interdisciplinary research (for example, this article). If, for example, middle authorships are seen by review committees as worthless (even, in some reported cases, detrimental) to a publication record, major contributors to the progress of science will go "extinct", unable to get their research funded. It is therefore important for biologists to learn to appreciate the value of many small contributions versus a few large contributions.

We hope to reach a day where quantitative skills are so pervasive (and valued) that calling someone a "computational biologist" sounds just as odd as a "pipette biologist" (Markowetz 2017). In the meantime, establishing separate structures and review schemes is another approach to support this class of researchers. To promote faculty success, many institutions have created Systems Biology, Computational Biology, or Biomedical Informatics departments to provide an environment in which researchers can thrive and be evaluated by like-minded interdisciplinary colleagues. Similarly, interdisciplinary journals, grant review panels, and funding schemes support the publication and funding of work evaluated by like-minded peers.

Likewise, institutions should take care to ensure that interdisciplinary researchers are recognized and rewarded for contributions across disciplines and departments, for example, through the evaluation system, additional compensation, and administrative staff. These researchers face more than their fair share of demands from collaborative roles on grants, administrative leadership, educational initiatives, thesis committees, and consultations. Many are jointly appointed to two or more departments, introducing additional service requirements that are invisible to each individual department (Yanai and Lercher 2020). Computational biologists can struggle to prioritize their research and these demands - even more so if they are in an under-represented demographic, as such scientists face disproportionate demands on their time and disproportionate costs (being labeled uncollaborative and unsupportive) if they decline.

**Develop and reward cross-disciplinary training and mentoring**
As large data sets become increasingly common, computational expertise is a valuable asset for any biologist. The deepest insights often result from data analyzed by the biologists who designed the experiment, and user-friendly software brings some analyses within reach of any researcher. Institutions can help researchers bridge gaps and gain computational skills.

The first step in acquiring practical bioinformatics skills is to become comfortable with computers and basic programming. Institutions can help here by providing general computer training courses that teach the command line, directory structure, and compiling and installing programs; this does not require computational biologists teachers who are in scarce supply. From there, institutions can support interdisciplinary trainees with educational programming and office hours tailored to training biologists in the basic principles of programming, data analysis, statistics, and machine learning. Teaching collaborative and interdisciplinary skills ideally begins at the undergraduate level where courses should be redesigned to blend computation, biology, and mathematics. Societies and other non-profits can play a major role here; examples include the Software Carpentries, CABANA in Latin America, and NEUBIAS in Europe.

Institutions can provide specific funding for interdisciplinary training programs and combine wet and dry lab training spaces where interactions among researchers with diverse expertise can flourish. As convenient as it may be, we caution against dry-lab researchers being physically segregated in separate buildings or sections of buildings from those with wet laboratories. For trainees who are embedded in laboratories outside their comfort zone, institutions and lab heads can provide mentors from their home domain and experiences that enable them to understand their unfamiliar domain, and external mentors should be rewarded institutionally for these contributions. We emphasize, however, that not all researchers need formal supervision and mentorship from a pure biologist or pure data scientist; many lead investigators in computational biology are quite experienced enough to cover both sides. A challenge of being interdisciplinary in many current academic structures is facing evaluation according to often conflicting, field-specific evaluation metrics. Mentors and departments should work to change evaluation systems to reward interdisciplinarity and, in the meanwhile, guide trainees to ensure their career goals can be met. In particular, oversight is needed for out-of-discipline trainees hired by individual PIs to be sure they are learning in their field and not treated as extremely cheap hired hands.

**Support computing and software development infrastructure to empower computational biologists**
Computing infrastructure (whether local cluster or cloud-based) is essential to modern biomedical data science and requires the combination of compute power, data storage, and networking - all of which can introduce significant costs to research projects. For most institutions, this computing infrastructure is often not centrally provided nor well-supported; it often falls through the cracks between institutional IT and research offices. As a result, individual laboratories often independently (and inefficiently) fund their own infrastructure and perform systems administration for clusters or the cloud that are far outside their actual expertise. Some funding agencies offer grant mechanisms that support computing hardware or cloud-based computing. Institutions that wish to foster computational research should subsidize infrastructure costs and provide support staff to assist in its use.

Quality development and maintenance of software is crucial for efficient and reproducible data analysis and is a key ingredient for successful computational biology projects (Levet et al. 2021). Community-level software ecosystems and pipeline-building tools have outsized impact because they standardize analyses and minimize software development costs for individual labs (Perkel 2021). Yet software maintenance and development is relatively undervalued in an academic system that prioritizes innovation, whether biological or computational. Software has a major impact on the progress of science but is underfunded by many agencies. The few agencies that do fund software maintenance are spread thin, given the global demand (Nowogrodzki 2019). Institutions and industry could meet this demand by providing centralized resources for software developers available part time to computational labs through fee for service models.

**Facilitate computationally-driven experimentation and data generation**
Historically, computational biologists relied on collaborators or public resources for datasets. It is now becoming more common for researchers with an informatics background to be running experiments, whether as a primary data source or to benchmark new technologies, validate algorithms, and test predictions and theories. Being able to bridge the dry lab-wet lab gap can have a major impact on a computational biologist's success. Institutions and laboratory neighbors can flexibly offer laboratory space for computational biologists when needed and can offer appropriate biology mentors for biologists joining computational groups. Core facilities can offer support: for example, if the typical service is to allow biologists to use an instrument, the facility can offer full service to computational laboratories for an additional fee.

**Provide incentives and mechanisms to share open data to empower discovery through reanalysis**
Data and code often become more valuable over time with new techniques or complementary new data. Indeed, some of the most challenging problems facing biological and clinical research are only possible through access to well-curated, large-scale datasets. Funding agencies, publishers, and the scientific community must continue to recognize both dissemination and reanalysis of reusable data as an impactful research output. FAIR (findable, accessible, interoperable, and reusable) principles have already been adopted and even required by many funding agencies and organizations (Wilkinson et al. 2016). New article types (e.g. Cell's "Resource") and entirely new journals (e.g. Gigascience, Scientific Data, and Data in Brief) provide a publication route, yielding academic credit for the creation, organization, and sharing of useful data sets.

A growing ecosystem of repositories is available for sharing various data types (Wilson et al., n.d.). Logistically, storing and disseminating data is not a trivial task even for those with some computational skill; it requires communication skills and knowledge encompassing information technology (IT), database structures, security/privacy, and desktop support that is distinct from analysis and the development of computational methods. Institutions and agencies might provide specialists in this and interfaces to ease the process. Creating mechanisms for capturing metadata and incentives that support high-quality annotations can help to return more value from computational analyses (Park and Greene 2018; Greene et al. 2017). To maximize the engagement of computational biologists, it is often helpful to provide data in a raw form as well as commonly used summary forms - controlling access as required to meet ethical and legal constraints (Byrd et al. 2020; Wilson et al., n.d.).

**Consider infrastructural, ethical, and cultural barriers to clinical data access**
Institutions with large medical centers are realizing the promise of discovery from multi-modal datasets of their unique patient cohorts, including primary clinical data as well as data from corresponding biospecimens using emerging molecular and imaging technologies. For example, deep learning has the potential to revolutionize pathology given sufficient data, often thousands of annotated images spanning patient groups. Infrastructural, ethical, and cultural considerations all place barriers to large-scale data access for computational research.

Institutions can play a role in supporting data access by providing centralized database structures that automatically ingest patient electronic health records and research-level data collected on these patients (Lau-Min et al. 2021). This requires careful attention to privacy, such that access to data is provided only to legally authorized researchers. It also requires policies in place governing whether patients will be notified about any potential health risks uncovered based on their data. Ethicists and clinical societies play a critical role in developing appropriate guidelines for such computational research on patient datasets that consider patient privacy and potential for improved public health. Inclusion of demographic information in these datasets and outreach to underrepresented populations is critical to overcome biases in data-driven discovery, and introduces further ethical considerations.

The current system rewards scientists and institutions that closely guard clinical datasets, because exclusive analysis can benefit careers and the data can be monetized due to commercial demand for novel biomarkers and therapeutic targets. Strong leadership is essential to incentivize academic investigators to collect datasets in a coordinated way across disease groups to enable research for public benefit. Federated learning, where machine learning models can be trained on multiple datasets without actually sharing the raw data, may work around some concerns.

**Conclusion**
As computational biology is a relatively new scientific discipline that has grown rapidly over the last 30 years, we may ask what computational biology will look like 30 years from now. The ever-growing amount of data and associated analytical questions will likely outstrip the supply of researchers with computational skills for the foreseeable future. On the other hand, as bioinformatics tools become more optimized and easy to use, more biologists will be able to analyze their data and may not need to rely on a dedicated computationalist. Likewise, as more data become open and contracted wet lab facilities grow, computationalists may not need to rely on dedicated wet lab scientists. Cultivating computational biology as described here is therefore crucial to fill this gap. One major demand for computational biology expertise will come from continued technological advances in molecular biology, providing novel types of data for which appropriate analysis methods have to be designed. This requires the development of a scientific culture that enables and promotes efficient collaboration across disciplines. Over time, the distinction between wet and dry biologists may fade, as both are working toward a common goal of understanding biology, and hybrid biologists (Eddy 2005;

Markowetz 2017) may emerge who are equally adept at the experimental and the computational aspects of molecular biology.

**Acknowledgments**
The authors thank the interdisciplinary researchers who have paved the way for their careers and the National Institutes of Health for research funding (R35 GM122547 to AEC, R01 HG010067 to CSG, and NCI U01CA253403 to EJF).